\documentclass{article}
\usepackage[export]{adjustbox}
\usepackage{putex}

\usepackage{cancel}
\usepackage{simplewick}
\usepackage{multirow}
\usepackage{comment}

\usepackage{graphicx}
\usepackage{caption}
\usepackage{amsmath}
\usepackage{amsfonts}
\usepackage{array}
\usepackage{subcaption}
\usepackage{epstopdf}
\usepackage{enumerate}
\usepackage{cite}
\usepackage{youngtab}
\usepackage{tensor}
\usepackage{slashed}
\usepackage[aligntableaux=center]{ytableau}
\usepackage{rotating}
\usepackage{bigfoot}

\usepackage[hidelinks]{hyperref}

\numberwithin{equation}{section}

\newcommand{\es}[2] {\begin{equation} \label{#1} \begin{split} #2 \end{split} \end{equation}}

\newcommand\thpnormind[5]{\langle {\cal O}_{\Delta_0}\left(#1\right) \hat{\cal O}_{\Delta_0}\left(#2\right) {\cal O}_{#3}^{#5}\left(#4\right) \rangle}
\newcommand\thpnorm[4]{\thpnormind{#1}{#2}{#3}{#4}{}}
\newcommand\thptildind[5]{\langle {\cal O}_{\tilde\Delta_0}\left(#1\right) \hat{\cal O}_{\tilde\Delta_0}\left(#2\right) {\cal O}_{#3}^{#5}\left(#4\right) \rangle}
\newcommand\thptild[4]{\thptildind{#1}{#2}{#3}{#4}{}}

\begin{document}

\institution{WZ}{Department of Particle Physics and Astrophysics, Weizmann Institute of Science, Rehovot, Israel}

\title{Explicit holography for vector models at finite $N$, volume and temperature}

\authors{
Ofer Aharony\worksat{\WZ}\footnote{e-mail: {\tt ofer.aharony@weizmann.ac.il}}, 
Shai M.~Chester\worksat{\WZ}\footnote{e-mail: {\tt iahs81@gmail.com}},
Tal ~Sheaffer\worksat{\WZ}\footnote{e-mail: {\tt sheaffertal@gmail.com}}
and Erez Y.~Urbach\worksat{\WZ}\footnote{e-mail: {\tt erez.urbach@weizmann.ac.il}}}

\abstract{
In previous work we constructed an explicit mapping between large $N$ vector models (free or critical) in $d$ dimensions and a non-local high-spin gravity theory on $AdS_{d+1}$, such that the gravitational theory reproduces the field theory correlation functions order by order in $1/N$. In this paper we discuss three aspects of this mapping. First, our original mapping was not valid non-perturbatively in $1/N$, since it did not include non-local correlations between the gravity fields which appear at finite $N$. We show that by using a bi-local $G-\Sigma$ type formalism similar to the one used in the SYK model, we can construct an exact mapping to the bulk that is valid also at finite $N$. The theory in the bulk contains additional auxiliary fields which implement the finite $N$ constraints. Second, we discuss the generalization of our mapping to the field theory on $S^d$, and in particular how the sphere free energy matches exactly between the two sides, and how the mapping can be consistently regularized. Finally, we discuss the field theory at finite temperature, and show that the low-temperature phase of the vector models can be mapped to a high-spin gravity theory on thermal AdS space.
}

\maketitle

\tableofcontents

\section{Introduction and Summary}

One of the simplest examples of the AdS/CFT correspondence \cite{Maldacena:1997re} is the duality between the $U(N)$-singlet sector of the theory of $N$ free complex scalar fields in $d$ space-time dimensions ($d > 2$) and a high-spin gravity theory on $AdS_{d+1}$ \cite{Klebanov:2002ja}. This example is free on the field theory side, and weakly coupled on the gravity side at large $N$, so in the large $N$ limit both sides are controllable (and this remains true also after deforming to the critical vector model). This suggests that at least order by order in $1/N$ it should be possible to explicitly map the field theory to the gravity side. Such an explicit mapping was presented in \cite{Aharony:2020omh} (following \cite{deMelloKoch:2018ivk}\footnote{Note that there are two different approaches to rewriting the vector model in terms of bi-local fields and mapping them to the bulk. One approach \cite{Koch:2010cy,Koch:2014aqa} uses the Hamiltonian language and bi-local operators at different points in space but at the same time; this breaks manifest Lorentz invariance and its relation to our approach is not clear. In particular, putting the theory at finite temperature is straightforward in the Hamiltonian approach, but as we discuss in section \ref{finiteT} it is non-trivial in our approach. Our approach, following \cite{deMelloKoch:2018ivk}, preserves Lorentz invariance (and, indeed, the full conformal group, giving strong constraints on the form of the mapping). In \cite{deMelloKoch:2018ivk} only the $d=2$ case was analyzed explicitly, and this case is more subtle because free scalar fields are not primary operators there, while \cite{Aharony:2020omh} derived an explicit mapping for $d > 2$.}); it rewrites the field theory as a non-local bulk theory of massless spin $J$ fields $\Phi_J$ on $AdS_{d+1}$, whose coupling constant is $1/N$. This mapping is derived by first changing variables in the field theory from the fields $\phi_I(x)$ ($I=1,\cdots,N$) to a $U(N)$-invariant bi-local field
\es{G}{
G(x_1,x_2) \equiv \frac{1}{N} \sum_{I=1}^N \phi_I^*(x_1) \phi_I(x_2)\,.
}
Then, the bi-local field is expanded in irreducible representations of the conformal group $SO(d+1,1)$ (we work in Euclidean space for simplicity). It was shown in \cite{Aharony:2020omh} that the expansion of massless transverse traceless spin $J$ fields $\Phi_J$ on $AdS_{d+1}$ (with $J=0,1,2,\cdots$) in representations of $SO(d+1,1)$ contains precisely the same representations, allowing a linear mapping (consistent with conformal invariance) between $G(x_1,x_2)$ and the bulk fields. It was conjectured that the resulting bulk action is equivalent in the classical limit to Vasiliev's high-spin gravity theory on $AdS_{d+1}$ \cite{Vasiliev:1990en,Vasiliev:1992av,Vasiliev:1995dn} (believed to be the unique consistent classical theory of high-spin fields on $AdS_{d+1}$), gauge-fixed such that it lives on a fixed $AdS_{d+1}$ background and such that the spin $J$ fields are transverse and traceless\footnote{It is not completely clear in the high-spin literature if Vasiliev's equations of motion lead to a consistent theory and also if this theory is local at long distances, see for instance the discussions in \cite{Boulanger:2015ova,Sleight:2017pcz}. In our work we do not use any specific properties of Vasiliev's theory; we construct an independent high-spin gravity theory directly from the CFT, and we believe that this should be equivalent to any other consistent formulation of a high-spin gravity theory. Whenever we refer to Vasiliev's theory we refer only to general properties of this theory which are independent of the unclear details. Note also that our bulk actions (reviewed below) have quartic-in-derivative kinetic terms, so they include additional modes beyond the physical transverse traceless fields of high-spin gravity. This agrees with the fact that the action for high-spin gravity cannot be fixed off-shell to include only transverse traceless fields (even though this can be done on-shell); we conjecture that the extra modes we have are related to the extra off-shell modes appearing in high-spin gravity theories, but this is still under investigation, and understanding this may be important for comparing our results to perturbative constructions of the high-spin gravity action along the lines of \cite{Bekaert:2016ezc,Bekaert:2014cea,Sleight:2016dba}. However, this does not play a role in this paper.}. Assuming that this is true, one may view the construction of \cite{Aharony:2020omh} as providing an action for this high-spin gravity theory, which enables for the first time both the computation of the action for classical solutions, and a systematic evaluation of quantum corrections.

The existence of an explicit off-shell mapping between field theory and bulk gravity path integrals, as constructed in \cite{Aharony:2020omh}, is highly non-trivial. The construction guarantees that any bulk computations of correlation functions on $\mathbb{R}^d$ will match with the free field theory perturbatively in $1/N$, but many aspects of the mapping are not yet clear.
In this paper we discuss three aspects and generalizations of the mapping of \cite{Aharony:2020omh}. 

In Section \ref{finiteN} we discuss how to perform the mapping for finite values of $N$. The mapping of \cite{Aharony:2020omh} was constructed such that it is valid order by order in the $1/N$ expansion. However, for finite $N$ the bi-local fields \eqref{G} obey constraints on products of $(N+1)$ such fields that were not taken into account in \cite{Aharony:2020omh}, and which translate into similar constraints on products of bulk fields that must arise at finite $N$. We can take this into account by introducing\footnote{We thank Douglas Stanford for suggesting this.} independent bi-local fields $G$ and $\Sigma$, such that $\Sigma$ is a Lagrange multiplier implementing the condition that $G$ is equal to \eqref{G}. This is similar to the standard method used, for instance, in the SYK model \cite{Maldacena:2016hyu} (for vector models this was previously done for instance in \cite{Benedetti:2018goh}, and the new thing we are doing is mapping this to the bulk). Integrating over the fields $\phi_I$ leads to an action for $G$ and $\Sigma$ which properly takes the constraints into account even at finite $N$. We can then map this action to the bulk, obtaining an extra set of bulk fields ${\tilde \Phi}_J$ which come from the auxiliary field $\Sigma$. The fields ${\tilde\Phi}_J$ may be viewed as auxiliary fields in the bulk, which implement also there the appropriate finite $N$ constraints; performing the path integral over them order by order in $1/N$ reproduces the bulk action of \cite{Aharony:2020omh}, but keeping them gives a consistent bulk theory even at finite $N$. Note that even at finite $N$ we can map the CFT to fields living on a fixed AdS background, and we do not need to introduce any sum over different geometries or topologies; presumably this is a special feature of high-spin gravity theories.

In Section \ref{finiteV}  we discuss the generalization of the mapping of \cite{Aharony:2020omh}, which was written for the field theory on $\mathbb{R}^d$, to a finite volume space, focusing on the sphere $S^d$. The generalization to $S^d$ is straightforward, since the theory on $S^d$ is related to the theory on $\mathbb{R}^d$ by a conformal transformation. However, understanding this generalization enables us to discuss two issues explicitly. First, the free energy on $S^d$ is finite (for finite $N$) and is one of the simplest observables in the vector models, so it is interesting to see how it is reproduced by the bulk gravitational theory. We find that in our finite $N$ formalism the sphere free energy is reproduced in a trivial way; it comes from a pre-factor in front of the bulk path integral, and does not receive any classical or quantum contributions from the bulk path integral. In the past, quantum corrections to naive versions of the bulk path integral of high-spin gravity (with specific off-shell continuations of Vasiliev's equations of motion) were compared to the field theory result \cite{Giombi:2013fka}, suggesting that for $O(N)$ theories the gravitational bulk coupling is actually $1/(N-1)$. We argue that in our case\footnote{The generalization of our results here for $U(N)$-singlets made from complex scalar fields to $O(N)$-singlets made from real scalar fields is straightforward.} the bulk computation is actually different than what was assumed in previous works, and it does not suggest any such shift. Second, explicit computations, both in the bi-local formalism and in the bulk, require regularization. We discuss for the theory on $S^d$ which regularizations are needed on both sides, and how we can perform the mapping between the two sides for the regulated theories in a way that ensures that a consistent mapping will arise when the regulators are taken to infinity. Additional details of this are described in an appendix.

Finally, in Section \ref{finiteT} we discuss the mapping to the bulk at finite temperature. Imposing the projection to $U(N)$ singlets at finite temperature is more complicated and requires introducing and integrating over a $U(N)$ holonomy, and we do not know how to map the field theory with the holonomy to the bulk. However, at large $N$, the integral over the holonomy is governed by a saddle point. In particular if we take the large $N$ limit before the large volume limit, the theory is in a low-temperature ``confining'' phase governed by a uniform distribution of the holonomy eigenvalues, while if we take the large volume limit first, the theory is in a ``deconfined'' phase with the holonomy matrix equal to the identity matrix. In the former case we show that at leading order in the saddle point approximation we can use the same mapping to the bulk as for zero temperature, and we obtain the same bulk theory but on a thermal AdS space, as expected. In the latter case we argue that the mapping must be modified, and presumably this corresponds to a bulk theory on some ``black hole'' background; we leave the understanding of this case to future work.

\section{Finite  \texorpdfstring{$N$}{Lg}}
\label{finiteN}

In this section we derive the AdS dual of a theory of $N$ free scalar fields at finite $N$. We start by reviewing how \cite{Jevicki:1979mb} mapped the path integral of $N$ free complex scalar fields\footnote{The results of this section can be easily extended to real scalar fields and $O(N)$-invariant bilocals, following the modifications discussed in \cite{Aharony:2020omh}.} $\phi_I(x)$ to a path integral of $U(N)$-invariant bilocals $G(x_1,x_2)$, which is only valid at large $N$ since at finite $N$ the bilocals obey constraints that were ignored. We then show how to make the bilocal path integral valid at finite $N$ by introducing a new bilocal field $\Sigma(x_1,x_2)$, and demonstrate how this path integral exhibits the expected finite $N$ constraints. We then describe how to map the finite $N$ bilocal theory to the bulk, as was done at large $N$ in \cite{Aharony:2020omh}, and show how the finite $N$ constraints set the amplitude of empty AdS to zero. Finally, we generalize to the critical $U(N)$ theory.

\subsection{Review of the bilocal theory at large \texorpdfstring{$N$}{Lg}}
\label{largeN}

We start with the local path integral for the free theory of $N$ complex scalars $\phi_I(x)$:
\es{localZ}{
Z &= \int D\phi_I(x) \exp \left( -S[\phi]  \right)\,,\qquad S[\phi] \equiv \sum_{I=1}^N \int d^d x | \vec \nabla \phi_I(x) | ^2\,,
}
which we would like to rewrite as a path integral in terms of the $U(N)$-invariant bilocal $G(x_1,x_2)$ defined in \eqref{G}. We will regulate the theory by placing it on a lattice of $V$ points, which provides both an IR and a UV regulator, so that $G(x_1,x_2)$ is a Hermitian $V\times V$ matrix. We then introduce $G(x_1,x_2)$ by inserting a delta function as
\es{ZtoG}{
Z &= \int DG(x_1,x_2) D\phi_I(x) \exp\left(-S[\phi] \right)\prod_{x_1,x_2} \delta\left(G(x_1,x_2)-\frac{1}{N}\sum_I \phi^{*}_{I}(x_1)\phi_{I}(x_2)\right) \\
&=\int DG(x_1,x_2) D\phi_I(x) \exp\left(-S[\phi]\right)\,\int D\tilde\Sigma(x_1,x_2)  \exp\left(2\pi i\text{Tr}\left(\tilde\Sigma\,G-\frac{1}{N} \sum_I \phi_{I}\tilde\Sigma\phi^{*}_{I}\right)\right)\,,
}
where in the second line we write the delta function as a path integral, and view $\phi_I^*(x_1) \phi_I(x_2)$ as a $V\times V$ matrix.\footnote{For the path integral to be finite we need to add a $-\epsilon\text{Tr}(\tilde\Sigma^2)$ for convergence, and take $\epsilon=0$ at the end.}  The continuum limit is reached by taking $V\to\infty$, in which case matrix traces become continuum integrals as ${\rm Tr}(G) \equiv \int d^d x G(x,x)$ and $(GH)(x_1,x_2) \equiv \int d^d x_3 G(x_1,x_3) H(x_3,x_2)$. We then use the delta function to set $S[\phi]=S_\text{free}[G]$, which we can write explicitly in the continuum limit as
\es{freeS2}{
 {S}_\text{free}[G]=N\int d^dx_1 \partial_{1,i} \partial_{2,i}G(x_1,x_2)\vert_{x_2=x_1}\,.
} 
The integrals over $\phi_I$ in \eqref{ZtoG} are then simple Gaussian integrals that give (up to an $N$-dependent constant):
\es{GSold}{
	Z = \int DG(x_1,x_2) \exp(-S_\text{free} [G]) \int D\tilde\Sigma(x_1,x_2) \exp\left(2\pi i  \text{Tr} (\tilde \Sigma  G)\right)
	\left|\det (\tilde \Sigma)\right|^{-N}.
}
Note that this step is not rigorous, since the integrals over $\phi_I$ diverge and require different analytic continuations depending on the sign of the eigenvalues of $\tilde\Sigma$, and the resulting integral over $\tilde\Sigma$ in \eqref{GSold} diverges near $\tilde\Sigma=0$; we will ignore these issues for now and fix them in the next subsection.

Finally, we change variables to $\Sigma' = \tilde \Sigma G$ and integrate over $\Sigma'$ to get (up to $N$ and $V$-dependent constants):
\es{eq:old_G_action}{
	Z = \int DG(x_1,x_2) \exp(-S_\text{free} [G]) \left|\det (G)\right|^{N-V}\,,
}
which explicitly depends on the regularization $V$. Note that this expression only converges if $V\leq N$, so if we want to consider the $V\to\infty$ continuum limit, then we must first take $N\to\infty$; this is related to finite-$N$ constraints on $G$ that we discuss below. Since the action is proportional to $N$, we can perform a large $N$ expansion around the saddle point of \eqref{eq:old_G_action} which is simply the free scalar propagator
\es{saddle}{
G_0(x_1,x_2)= \frac{\Gamma(d/2-1)}{4\pi^{d/2}} |x_{12}|^{2-d}\,,
}
where $x_{12}\equiv |x_1-x_2|$, which yields the expected $V$-independent answer for correlators in the free theory \cite{Jevicki:1979mb,Aharony:2020omh}. Note that the bilocal action is interacting, so to get the free theory result one must cancel different nonzero contributions at each order in $1/N$.

\subsection{Bilocal theory at finite \texorpdfstring{$N$}{Lg}}
\label{finiteN2}

To define the continuum bilocal action at finite $N$, we must take $V\to\infty$ at finite $N$, so the approach above is not suitable, and we will describe a different approach here. We start with the continuum version of the path integral in \eqref{ZtoG} over $G(x_1,x_2)$, $\tilde\Sigma(x_1,x_2)$, and $\phi_I(x)$. We change variables in the path integral to
\es{newFields}{
 \eta(x_1,x_2) = N(G(x_1,x_2)-G_0(x_1,x_2))\,,\qquad \Sigma(x_1,x_2) = \frac{1}{N} \tilde\Sigma(x_1,x_2)\,,
}
which has a trivial Jacobian.\footnote{In the large $N$ approach in \cite{Aharony:2020omh}, we defined $\eta(x_1,x_2)$ as a fluctuation around the large $N$ saddle $G_0(x_1,x_2)$, where we rescaled $\eta(x_1,x_2)$ by $\sqrt{N}$ for convenience in the large $N$ expansion. Here, we use a different normalization.} We then perform the Gaussian integral over $\phi_I(x)$, this time keeping $S[\phi]$ as a function of $\phi_I(x)$ so that the Gaussian path integral over $\phi_I$ is convergent, to get the path integral
\es{eq:eta_sigma_Z}{
	Z= \int D\eta(x_1,x_2) \int D\Sigma(x_1,x_2) 
	\exp\left( -S[\eta,\Sigma] \right)\,,\qquad S[\eta,\Sigma] = -2\pi i \text{Tr}(\Sigma\eta) - N  W[\Sigma]\,,
}
where the integral over $\phi$ gives (using the fact that $G_0$ is minus the inverse of the Laplacian)
\es{eq:W_def}{
	W[\Sigma] &= \frac{1}{N}\log \left[ \int D\phi_I(x) 
	\exp\left(-\int d^d x \sum_{I=1}^N \partial_i \phi_I^*(x) \partial_i \phi_I(x)-2\pi i \text{Tr}\left( \Sigma  \tilde\eta[\phi_I] \right) \right) \right] \\
& = \text{Tr}\left(
	2\pi i  G_0\Sigma \right) - \log(\det[-(\square - 2 \pi i \Sigma)/\pi]) \\
	& = \log (Z_\phi) + \text{Tr}\left(
	2\pi i  G_0\Sigma - \log \left(1+2\pi i G_0  \Sigma\right)\right),
}
where $\tilde\eta[\phi_I](x_1,x_2) \equiv \sum_I \phi_I^*(x_1) \phi_I(x_2) - N G_0(x_1,x_2)$, 
and we define the usual free scalar partition function
\begin{equation}\label{eq:Z_phi}
	Z_\phi = \int D\phi(x) \exp\left(-\int d^d x \partial_i \phi^*(x) \partial_i \phi(x)\right) = \det(\pi G_0)\,,
\end{equation}
which is divergent in the continuum limit, and so will depend explicitly on $V$. Note that unlike the large $N$ action in \eqref{eq:old_G_action}, the $V$ dependence here is just in an overall factor multiplying the path integral. 

To derive the Feynman rules for this path integral, we expand the action \eqref{eq:eta_sigma_Z} to quadratic order in $\Sigma$ and $\eta$ to get
\es{eq:S2_bi}{
	& S^{(2)}[\eta,\Sigma] = \int d^d x_1 d^d x_2 d^d x_3 d^d x_4 \\
	& \times \frac{1}{2} \left[\begin{matrix}\eta(x_1,x_2) & 2\pi \Sigma(x_1,x_2)\end{matrix}\right]
	\left[\begin{matrix}
		0 & -i \delta(x_1,x_4)\delta(x_2,x_3)\\
		-i\delta(x_1,x_4)\delta(x_2,x_3) &  
		N G_0(x_1,x_4) G_0(x_2,x_3)
	\end{matrix}\right]
	\left[\begin{matrix}\eta(x_3,x_4)\\
2\pi\Sigma(x_3,x_4)
\end{matrix}\right]\,,
}
which yields the contractions
\es{eq:bilocal_contrac}{
	\contraction{}{\eta(x_1,x_2)}{}{2\pi \Sigma(x_3,x_4)}
	\eta(x_1,x_2) 2\pi \Sigma(x_3,x_4) &= 
	\includegraphics[valign=c]{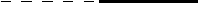}
	=i \delta(x_1-x_4)\delta(x_2-x_3),\\
	\contraction{}{\eta(x_1,x_2)}{}{\eta(x_3,x_4)}
	\eta(x_1,x_2) \eta(x_3,x_4) &= 
	\includegraphics[valign=c]{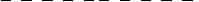}
	=N G_0(x_1,x_4)G_0(x_2,x_3)\,.
}
Higher order vertices come only from $W[\Sigma]$ and include the $\Sigma$-only $n$-point vertices ($n \geq 3$)
\begin{equation}
	\prod_{i=1}^n(2\pi \Sigma(x_i,y_i)) \text{ vertex} 
	=
	\includegraphics[valign=c]{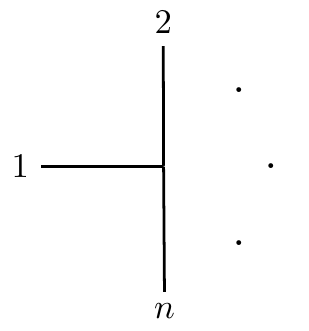}
	= N  \frac{(-i)^n}{n} G_0(y_1,x_2) G_0(y_2,x_3)\cdots G_0(y_n,x_1)\,.
\end{equation}
Since there are no $\Sigma$ self-contractions, the theory has no loop diagrams. Up to permutations, there is a single connected diagram for $\langle \eta^n \rangle$ with $n>2$, which comes from contracting it with the $\Sigma^n$ vertex. As a result, the theory requires no further regularization (or counter-terms) and is well-defined for finite $N$, manifestly giving the same answers as the free field contractions. This contrasts with the large $N$ path integral discussed in the previous subsection, whose diagramatic expansion requires counter-terms and nontrivial cancellations at each order in $1/N$ to get the expected cancellation of these corrections. Since the $1/N$ expansion of free theory correlators is exact, there is ultimately no difference between the large $N$ and finite $N$ formulations for these observables. 

Note that by construction, if we integrate out $\Sigma$ order by order in perturbation theory in $1/N$ we get back the action of Section \ref{largeN}. However, non-perturbatively it is not clear how to perform this integral over $\Sigma$.

The main difference between the large $N$ path integral of Section \ref{largeN} and the finite $N$ path integral \eqref{eq:eta_sigma_Z} is that only the latter exhibits the non-linear off-shell constraints between products of $N+1$ bilocals $G(x_1,x_2)$, which come from the definition in \eqref{G} in terms of $\phi_I(x)$; this is a consistency check on our finite $N$ formalism. For instance, for $N=1$ we expect the constraints
\es{N1}{
N=1:\qquad G(x_1,x_2) G(x_3,x_4) = G(x_1,x_4) G(x_3,x_2)\,.
}
 For general $N$, we can consider the points $x_1,...,x_{N+1}$ and $y_1,...,y_{N+1}$, and then the constraint on the $(N+1)\times(N+1)$ matrix $ {\hat G}_{ij}\equiv G(x_i,y_j)$ is that it has determinant zero:
\begin{equation}\label{eq:constraint}
	\det  ({\hat G}) = \sum_{\sigma\in S_{N+1}} (-1)^\sigma G(x_1,y_{\sigma(1)})...G(x_{N+1},y_{\sigma(N+1)})=0\,,
\end{equation}
which follows from the definition \eqref{G}, and reduces to \eqref{N1} when $N=1$. We can also consider the constraints using the finite $V$ regularization, in which case $ G(x_1,x_2)$ is a $V\times V$ matrix that must satisfy the constraint
\begin{equation}\label{eq:rank}
	\text{rank}\;  (G) \le N\,.
\end{equation}
This follows from the fact that we can write $N$ functions $\phi_I(y)$ as linear combinations of $N$ rows $G(x_I,y)$ for all $y$ and some $x_1,\dots,x_N$, so that any other row $G(x,y)$ can be written as a  linear combinations of the other $N$ rows $G(x_I,y)$. When all the $\phi_I$'s are different, we saturate the bound $\text{rank}\; (G) = N$. These constraints are not taken into account in the large $N$ path integral \eqref{eq:old_G_action}. 

To see these constraints in the finite $N$ path integral \eqref{eq:eta_sigma_Z}, we integrate over $\Sigma$, which localizes the $\eta$ integration measure to configurations that satisfy \eqref{eq:constraint} (or equivalently \eqref{eq:rank}). In more detail, let $f[\eta]$ be a functional of $\eta(x_1,x_2)$. In the $\Sigma$ path integral \eqref{eq:eta_sigma_Z} $\eta$ couples only linearly to $\Sigma$, and $\Sigma$ couples inside \eqref{eq:W_def} only linearly to $\tilde \eta$. As a result, integration by parts gives
\begin{equation}
	\int D\Sigma(x_1,x_2) f[\eta]
	\exp\left( -S[\eta,\Sigma] \right) 
	= Z_\phi^N  
	\int D\Sigma(x_1,x_2) e^{2\pi i \text{Tr}( \Sigma  \eta)} 
	\langle f[\tilde\eta[\phi_I]] e^{-2\pi i \text{Tr}( \Sigma  \tilde\eta[\phi_I]) }\rangle_{\phi}\,,
\end{equation}
where $\langle\rangle_\phi$ denotes the expectation value in the local path integral \eqref{localZ}. Now take $f[\eta]=f[N(G-G_0)]$ to be any of the constraints \eqref{eq:constraint}. Since $f[ \eta]=0$ vanishes as an operator in the local theory, the right-hand side must also vanish inside the path integral. 

\subsection{AdS/CFT map}
\label{adscftmap}

We will now write down the finite $N$ map between these bilocal fields and bulk fields living on a fixed anti-de Sitter background for every integer spin $J$. For $\eta$, the map is the same as in \cite{Aharony:2020omh}, which we briefly review now. We start by expanding $\eta( x_1 , x_2)$ in the complete basis \cite{Dobrev:1977qv}
\es{eta_decomp}{
	\eta(x_1,x_2) = \sum_{J=0}^\infty \int_{\gamma_J}\frac{d\Delta}{2\pi i} \int d^d y  \, C_{\Delta,J}\left(y\right)
	\thpnorm{x_1}{x_2}{\Delta,J}{y}\,,
}
where the basis elements are ``3-point functions''\footnote{Note that while the harmonic basis resembles a three point function, it does not correspond to a correlator in a physical CFT, and is simply a useful basis for conformally-covariant functions.} of scalar operators $\mathcal{O}_{\Delta_0}$ and $\hat{\mathcal{O}}_{\Delta_0}$ that have the same scaling dimension $\Delta_0=\frac{d-2}{2}$ as a free scalar field and an operator $\mathcal{O}_{\Delta,J}$ of scaling dimension $\Delta$ and spin $J$, namely
\es{3point}{
	\thpnormind{x_1}{x_2}{\Delta,J}{x_3}{i_1,...,i_J} = \frac
	{Z^{i_1}...Z^{i_J}-\text{traces}}
	{x_{12}^{2\Delta_{0}-\Delta+J}x_{13}^{\Delta-J}x_{23}^{\Delta-J}}\,,
}
where $Z^{i} \equiv \frac{x_{13}^i}{x_{13}^2}-\frac{x_{23}^i}{x_{23}^2}$, $x_{ij} \equiv x_i-x_j$, and we will in general suppress spin indices for simplicity. The contours $\gamma_J$ of the $\Delta$ integrals in \eqref{eta_decomp} go over the principal series $\Delta=\frac d2+is$ for real $s$, except for $J=0$ and $d<4$ where we deform the contour \cite{Aharony:2020omh} to ensure that the pole at $\Delta=2\Delta_0=d-2$ of the scalar single-trace operator in the free theory appears on the same side of the contour as the other spin $J$ single-trace operators with scaling dimension $d-2+J$. The expansion \eqref{eta_decomp} exists when $\eta(x_1,x_2)$ is Hermitian ($\eta^*(x_1,x_2) = \eta(x_2,x_1)$) and satisfies the conditions \cite{Dobrev:1977qv}
\begin{enumerate}
	\item $\lim_{x_2\to x_1}\eta(x_1,x_2)$ should be finite.
	\item At large $|x_1+x_2|$ (and fixed difference) $\eta(x_1,x_2)$ should decay.\footnote{In \cite{Aharony:2020omh}, a related map was also derived for $\eta(x_1,x_2)$ that does not satisfy this condition.}
	\item At large $|x_1|$ and fixed $x_2$
	\begin{equation}
		\eta(x_1,x_2) \sim \left|x_{1}\right|^{-2\Delta_{0}}\cdot \text{Power series in }\ensuremath{\frac{1}{|x_1|}}\,. \label{eq:bc_eta}
	\end{equation}
	\item $\eta(x_1,x_2)$ must be smooth.
\end{enumerate}

For $d>4$, we can use the orthogonality and completeness relations of the 3-point basis, as reviewed in Appendix A of \cite{Aharony:2021ovo}, to invert \eqref{eta_decomp} and write $C_{\Delta,J}(y)$ in terms of $\eta(x_1,x_2)$ as
\es{coeff}{
d>4:\qquad	C_{\Delta,J}(y)=\frac12\frac{1}{N_{\Delta,J}}\int d^dx_{1}d^dx_{2}\, \eta\left(x_{1},x_{2}\right)
	\thptild{x_1}{x_2}{\tilde\Delta,J}{y}\,,
}
where the normalization $N_{\Delta,J}$ is
\es{CFTNorm}{
	N_{\Delta,J}=\frac{\pi^{\frac{3d}{2}}\Gamma(J+1)}
	{2^{J-1}\Gamma(\frac{d}{2}+J)}
	\frac{\Gamma\left(\Delta-\frac{d}{2}\right)}{\Gamma\left(\Delta-1\right)\left(\Delta+J-1\right)}\frac{\Gamma\left(\tilde{\Delta}-\frac{d}{2}\right)}{\Gamma\left(\tilde{\Delta}-1\right)\left(\tilde{\Delta}+J-1\right)}\,,
	}
	and tildes over dimensions denote $\tilde\Delta\equiv d-\Delta$.

	We then consider spin $J$ transverse traceless bulk fields $\Phi_J(x,z)$ defined on AdS$_{d+1}$ using the Poincar\'e coordinates
	\es{AdSmet}{
	ds^2 = \frac{dx^i dx^i + dz^2}{z^2}\,.
	}
	We expand these fields in the complete basis of bulk-to-boundary propagators
\es{bulk_comp}{
\Phi_{J}(x,z)=\int_{\gamma_J}\frac{d\Delta}{2\pi i}\int 
	d^dy
	f_{\Delta,J}C_{\Delta,J}(y)G_{\Delta,J}(x,z|y)\,,
	}
	and identify the bulk coefficient $C_{\Delta,J}(y)$ with the same coefficient appearing in the bi-local expansion \eqref{eta_decomp}, up to a multiplicative factor $f_{\Delta,J}$ that is not fixed by conformal symmetry (it is related to the freedom of performing field redefinitions in the bulk). This identification implies that the contour $\gamma_J$ in \eqref{bulk_comp} is the same as the one in the bi-local expansion. The basis elements in \eqref{bulk_comp} are the bulk-to-boundary propagators in AdS space, defined by the differential equation
\es{eq:bulk_boundary_DE}{
	&(\nabla^2_{x,z} -M^2_{\Delta,J})G _{\Delta,J}(x,z|y)=  0\,,\qquad \nabla^2_{x,z} \equiv z^{d+1}\partial_{z}\left(z^{-d+1}\partial_{z}\right)+z^{2}\nabla_x^2\,,\\
}
where $M^2_{\Delta,J} \equiv \Delta(\Delta-d) - J$, and by the $z\to0$ boundary condition
 \es{eq:2_point_limit}{
  	G_{\Delta,J}(x,z|y)=z^{\Delta-J} 
 	\langle\mathcal{O}_{\Delta,J}(x)\mathcal{O}_{\Delta,J}(y)\rangle + z^{d-\Delta-J} S_B ^{\Delta,J} \delta_J^{TT}(x-y)+ \dots\,.
}
Here $\delta_J^{TT}(x)$ denotes a delta function with $2J$ suppressed lower indices for spin $J$ traceless transverse functions on the boundary $\mathbb{R}^d$, and we define the bulk shadow coefficient
\es{eq:bulk_shadow_coeff}{
	S_B^{ \Delta,J} 	\equiv \frac{\pi^{\frac{d}{2}} \Gamma( \Delta-\frac{d}{2})}
	{(J+ \Delta-1)\Gamma(\Delta-1)}\,.
}
The expansion \eqref{bulk_comp} assumes that $\Phi_J(x,z)$ decays in the small $z$ limit as $z^{\frac d2-J}$ for $d>4$, and $\Phi_J(x,z)$ should also decay at large $x$, which corresponds to $\eta(x_1,x_2)$ decaying at large $|x_1+x_2|$.\footnote{A map for non-decaying $\Phi_0$ was also derived in \cite{Aharony:2020omh}.} The CFT-to-AdS map and its inverse now simply follow from a convolution of the bulk and CFT bases, and takes the explicit form
\es{CFTtoAdS}{
d>4:\quad	\Phi_{J}(x,z)& =\frac12
	\int_{\gamma_J}\frac{d\Delta}{2\pi i}\frac{f_{\Delta,J}}{N_{\Delta,J}}\int d^dy\int d^dx_1 d^dx_2  \, G_{\Delta,J}(x,z|y)
\thptild{x_1}{x_2}{\tilde\Delta,J}{y}\eta(x_1,x_2)\,,\\
	\eta(x_1,x_2) & = 
			\sum_{J=0}^\infty 
			\int_{\gamma_J}\frac{d\Delta}{2\pi i} 
			\int d^dy
			\int \frac{d^dxdz}{z^{d+1}} 
			\frac{G_{\tilde{\Delta},J}(x,z|y)}{\alpha_J \, N_{\Delta,J} f_{\Delta,J}}
			\thpnorm{x_1}{x_2}{\Delta,J}{y}
			\Phi_{J}(x,z)\,,
}
where $\alpha_J$ comes from the bulk orthogonality relation and has the value
\es{eq:alpha_J_def}{
 	\alpha_J  \equiv \frac{2^J \Gamma(\frac{d}{2}+J)}{\pi^{\frac{d}{2}}\Gamma(J+1)}\,.
 	} 
For $d<4$, the expansion of the bilocal in \eqref{eta_decomp} does not converge, so we must define an auxiliary bilocal and derive the map in terms of that. As shown in \cite{Aharony:2020omh}, this ultimately gives the modified CFT-to-AdS map
\es{ds4}{
d<4:\qquad	\Phi_{J}(x,z)& = \frac12 \int_{\gamma_J} \frac{d\Delta}{2\pi i}  \int d^dy \int d^dx_1 d^dx_2 \frac{f_{\Delta,J}}{\lambda_{\Delta,J}N_{\Delta,J}} \\
		&\quad \times  
		G_{\Delta,J}(x,z|y)
		\thpnorm{x_1}{x_2}{\tilde\Delta,J}{y} \, \nabla^2_1\nabla^2_2\eta\left(x_1,x_2\right)\,,
}
while the AdS-to-CFT map is the same as the $d>4$ case in \eqref{CFTtoAdS} except that we need to replace the $J=0$ term by
\es{J0}{
			\int_{\gamma_0}\frac{d\Delta}{2\pi i} 
			\int d^dy \int \frac{d^dxdz}{z^{d+1}}  
			\thpnorm{x_1}{x_2}{\Delta,0}{y}
			 \frac{G_{\tilde{\Delta},0}\left(x,z|y\right)}{\alpha_0 N_{\Delta,0} \lambda_{\Delta,0} f_{\Delta,0}}\\
			\times 
			\left(\nabla_{x,z}^{2}-M_{d-2,0}^{2}\right)\left(\nabla_{x,z}^{2}-M_{d,0}^{2}\right)\Phi_{0}(x,z)\,.
}
Here $\lambda_{\Delta,J}$ is the eigenvalue of the bi-local Laplacian in the conformal basis and is given for any $d$ by
\es{lam}{
\lambda_{\Delta,J}=\left(M_{\Delta,J}^{2}-M_{d+J,J}^{2}\right)\left(M_{\Delta,J}^{2}-M_{d+J-2,J}^{2}\right)\,.
} 

We can similarly expand $\Sigma(x_1,x_2)$ in terms of another set of bulk fields $\tilde\Phi_J$. From the term $\text{Tr}(\eta\Sigma)$ in the bilocal action \eqref{eq:eta_sigma_Z}, we see that $\Sigma$ transforms under conformal transformations as a product of two $\tilde\Delta_0=\frac{d+2}{2}$ scalar primaries, and so can be expanded in the complete basis\footnote{As a check, we can compute the two point function of $\eta$ using the conformal basis of the $C_{\Delta,J}$ and $\tilde C_{\Delta,J}$, as was done in Appendix B of \cite{Aharony:2020omh} for the large $N$ path integral in terms of just $C_{\Delta,J}$. In the conformal basis, the $\eta,\Sigma$ paring becomes
\begin{equation}
	\text{Tr}\left( \Sigma  \eta\right) = \sum_{J=0}^\infty \int_{\gamma_J} \frac{d\Delta}{2\pi i} \int d^d y
	\; N_{\Delta,J} \; C_{\Delta,J}(y)  \tilde C_{\Delta,J}(y).
\end{equation}
This implies that the analysis with only $C_{\Delta,J}$ will generalize to the $C_{\Delta,J},\tilde C_{\Delta,J}$ case as long as the on-shell result is spanned by the products $\thpnorm{x_1}{x_2}{\Delta,J}{y} \thpnorm{x_3}{x_4}{\Delta',J'}{y'}$, as we expect.}\footnote{The $\tilde C_{\Delta,J}$ here are not to be confused with the same notation we used in \cite{Aharony:2020omh} for the $d<4$ theory.}
\es{eq:sigma_conf_basis}{
	\Sigma(x_1,x_2) &= 
	\sum_{J=0}^\infty \int_{\gamma_J} \frac{d\Delta}{2\pi i} \int d^d y
	\;  \tilde C_{\Delta,J}(y)\;
	\thptild{x_1}{x_2}{\tilde\Delta,J}{y}\,,
}
which can be thought of as a ``shadow version'' of \eqref{eta_decomp}, and the contour $\gamma_J$ is the same as there. For this expansion to converge, we require the same boundary conditions for $\Sigma(x_1,x_2)$ that we gave for $\eta(x_1,x_2)$ in \eqref{eq:bc_eta}, except that we modify the third condition to 
\es{eq:bc_Sig}{
	\Sigma(x_1,x_2) \sim \frac{1}{|x_1|^{d+2}}  \cdot \text{Power series in }\frac{1}{|x_1|}\,.
}
We then expand a new set of spin $J$ transverse traceless bulk fields $\tilde\Phi_J(x,z)$
\es{bulk_comp2}{
\tilde\Phi_{J}(x,z)=\int_{\gamma_J}\frac{d\Delta}{2\pi i}\int 
	d^dy
	{\tilde f}_{\Delta,J}\tilde C_{\Delta,J}(y)G_{\Delta,J}(x,z|y)\,,
	}
where ${\tilde f}_{\Delta,J}$ parameterize bulk field redefinitions, analogous to ${ f}_{\Delta,J}$ in \eqref{bulk_comp}. We can then convolve \eqref{eq:bc_Sig} with \eqref{bulk_comp2} to get the map
\es{CFTtoAdS2}{
d>4:\qquad		\Sigma(x_1,x_2) &= 
	\sum_{J=0}^\infty \int_{\gamma_J} \frac{d\Delta}{2\pi i} \int d^d y \int \frac{d^d x dz}{z^{d+1}}   \frac{G_{\Delta,J}(x,z|y)  }{ \alpha_J N_{\Delta,J} \tilde f_{\tilde\Delta,J}} 
	\thptild{x_1}{x_2}{\tilde\Delta,J}{y}
	\; \tilde\Phi_J(x,z)\,,
}
where for $d<4$ we again replace the $J=0$ term by 
\es{J02}{
			\int_{\gamma_0}\frac{d\Delta}{2\pi i} 
			\int d^dy \int \frac{d^dxdz}{z^{d+1}}  
			\thptild{x_1}{x_2}{\tilde\Delta,0}{y}
			 \frac{ G_{{\Delta},0}\left(x,z|y\right)}{\alpha_J N_{\Delta,0}  \tilde f_{\tilde\Delta,0}  \lambda_{\Delta,0}}\\
			\times 
			\left(\nabla_{x,z}^{2}-M_{d-2,0}^{2}\right)\left(\nabla_{x,z}^{2}-M_{d,0}^{2}\right)\tilde\Phi_{0}(x,z)\,.
}
Note that we replaced $\Delta\to\tilde\Delta$ relative to the $\eta$ map \eqref{CFTtoAdS} for later convenience.

\subsection{Bulk dual at finite \texorpdfstring{$N$}{Lg}}
\label{bulkAction}

We can now use the AdS/CFT maps for $\eta$ and $\Sigma$ to derive the finite $N$ bulk action from the bilocal action \eqref{eq:eta_sigma_Z}. In the large $N$ case considered in \cite{Aharony:2020omh}, we chose the specific bulk field redefinition 
\es{flocal}{
f^\text{local}_{\Delta,J} = \left( (-1)^J \frac{S^{\tilde\Delta,J}_B}{S^{\tilde\Delta,J}_{\Delta_0,\Delta_0}}\right)^{\frac{1}{2}}
}
to get a local quadratic term in the bulk. In the finite $N$ case we consider now, we first choose ${\tilde f}_{\Delta,J}$ to be
\es{fnew}{
{\tilde f}_{\Delta,J}=\frac{2}{\alpha_Jf_{\tilde\Delta,J}}\,,
}
which for any $f_{\Delta,J}$ makes the bulk version of $\text{Tr}(\eta\Sigma)$ local:
\begin{equation}\label{eq:local_coupling}
	\int dx_1 dx_2 \Sigma(x_2,x_1) \eta(x_1,x_2) = \sum_{J=0}^\infty \int \frac{d^d x dz}{z^{d+1}} \tilde\Phi_J(x,z) \Phi_J(x,z)\,,
\end{equation}
 as follows from the completeness and orthogonality relations in Appendix A of \cite{Aharony:2021ovo}. Another advantage of this choice is that the Jacobian of the combined change of variables from the AdS/CFT maps \eqref{CFTtoAdS} and \eqref{CFTtoAdS2} is $1$, so that the partition function \eqref{eq:eta_sigma_Z} (without any overall factors) can be written as
\es{eq:bulk_Z}{
	Z_\text{bulk} = \int D\Phi_J(x,z) D\tilde\Phi_J(x,z) \exp\Big( 2\pi i  \sum_{J=0}^\infty \int \frac{d^d x dz}{z^{d+1}} \tilde\Phi_J(x,z) \Phi_J(x,z) + N  W[\tilde\Phi_J]\Big)\,,
}
where $W[\tilde\Phi_J]$ is a non-local expression given by substituting \eqref{CFTtoAdS2} into $W[\Sigma]$ in \eqref{eq:W_def}. In \eqref{eq:bulk_Z}, $N$ can be any real positive number\footnote{For non-integer values of $N$ the theory may be non-unitary, as in \cite{Maldacena:2011jn}.} and is identified with the inverse gravitational constant $G_N=1/N$ of the higher-spin gravity theory. We can then choose $f_{\Delta,J}=f^\text{local}_{\Delta,J} $ so that we obtain a local bulk action up to quadratic order in the fields,
\es{eq:S2_bi2}{
	& S^{(2)}[\Phi_J,\tilde\Phi_J] = \sum_{J=0}^\infty \int \frac{d^d x dz}{z^{d+1}}  \\
	& \times \frac{1}{2} \left[\begin{matrix} \Phi_J(x,z)& 2\pi  \tilde\Phi_J(x,z) \end{matrix}\right]
	\left[\begin{matrix}
		0 &-i \\
		-i&  
		\frac{\alpha_J N}{2}  \frac{ 
				\Pi^{TT}_{d-2+J,J}(x_1,z_1|x_2,z_2)-\Pi^{TT}_{d+J,J}(x_1,z_1|x_2,z_2)
			}{M^2_{d+J,J}-M^2_{d+J-2,J}} 
	\end{matrix}\right]
	\left[\begin{matrix} \Phi_J(x,z)\\
2\pi  \tilde\Phi_J(x,z)
\end{matrix}\right]\,,
}
which gives a local propagator in the bulk Feynman rules
\es{eq:bulk_contrac}{
&\contraction{}{\Phi_{J_1}}{(x_1,z_1)2\pi}{\tilde\Phi_{J_2}}
	\Phi_{J_1}(x_1,z_1) 2\pi \tilde\Phi_{J_2}(x_2,z_2) =
	\includegraphics[valign=c]{figures/eta_sigma.pdf} = 
	i \; \delta_{J_1,J_2} z_1^{d+1}\delta^{(d)}(x_1-x_2)\delta(z_1-z_2),\\
\contraction{}{\Phi_{J}}{(x_1,z_1)}{\Phi_{J}}
	&\Phi_{J}(x_1,z_1) \Phi_{J}(x_2,z_2) = \includegraphics[valign=c]{figures/eta_eta.pdf}\\
	&\quad = \frac{\alpha_J N}{2}  \frac{ 
				\Pi^{TT}_{d-2+J,J}(x_1,z_1|x_2,z_2)-\Pi^{TT}_{d+J,J}(x_1,z_1|x_2,z_2)
			}{M^2_{d+J,J}-M^2_{d+J-2,J}} \,.
}
Here, $\Pi^{TT}_{d-2+J,J}(x_1,z_1|x_2,z_2)$ is the traceless transverse bulk-to-bulk propagator defined in \cite{Aharony:2020omh} by the differential equation
\es{eq:bulk_bulk_DE_2}{
 	\left( \nabla^2_{x,z} - M^2_{\Delta,J}\right) \Pi^{TT}_{\Delta,J}(x_1,z_1|x_2,z_2){}   = -\delta^{TT}(x_1,z_1|x_2,z_2)\,,
}
and by the $z_2\to0$ boundary condition
\es{eq:BB_limit}{
	\Pi^{TT}_{\Delta,J}(x_1,z_1|x_2,z_2) =
	\begin{cases}
{z_2}^{\Delta-J} {\cal C}_{\Delta,J} G_{\Delta,J}(x_1,z_1|x_2) +O(z_2^{\Delta-J+1}) & \Delta<d-J+4\\
O(z_2^{d-2J+4})& \Delta\geq d-J+4\,, \\
\end{cases}
}
where $\delta^{TT}$ is a delta function for traceless transverse functions that is defined in \cite{Aharony:2020omh}, and the normalization
is 
\es{eq:bb_limit_coeff}{
	{\cal C}_{\Delta,J} \equiv \frac{(J+\Delta-1)\Gamma(\Delta-1)}{2\pi^{\frac{d}{2}} \Gamma(\Delta+1-\frac{d}{2})}\,.
}
The rest of the vertices come from $W[\tilde\Phi]$. Since these are $\tilde\Phi^n$ vertices ($n\geq 3$), there are no bulk loop diagrams and no need for counter-terms. Up to permutations there is a single (tree level) connected diagram that contributes to $\langle\Phi_{J_1} \cdots \Phi_{J_n}\rangle$, which gives exactly the mapping of the free $U(N)$ result.

As in Section \ref{finiteN2}, order by order in $1/N$ we can integrate out $\tilde\Phi$ to obtain an action for $\Phi$ which is the same as the action constructed in \cite{Aharony:2020omh}; this action has the same physical field content as Vasiliev's higher spin gravity theory and it was conjectured to be a gauge-fixed form of this theory. However, to define the theory non-perturbatively in $N$, it seems that the auxiliary field $\tilde\Phi$ is necessary.

We can demonstrate the utility of a finite $N$ bulk action by computing the simplest observable that is expected to vanish due to finite $N$ constraints. In the CFT, 
when $N < V$, one constraint that follows from \eqref{eq:rank} is that $G$ cannot equal $G_0$, or $\eta$ cannot vanish.
We saw in Section \ref{finiteN2} that the $\eta$ measure, after integrating over $\Sigma$, is localized to $\eta$'s that satisfy the finite $N$ constraints. In other words, if we denote this measure by
\begin{equation}
	\mu[\eta] = \int D\Sigma(x_1,x_2) e^{-S[\eta,\Sigma]}\,,
\end{equation}
then the claim is that $\mu[\eta = 0]=0$ for $N<V$. Since the Jacobian of the AdS/CFT map is 1, this translates in the bulk to
\begin{equation}
	\mu_\text{bulk}[\Phi_J] \equiv \int D\tilde\Phi_J e^{-S[\Phi,\tilde\Phi]} = \mu[\eta(\Phi_J)]\,,
\end{equation}
where $\eta(\Phi_J)$ is given in terms of the bulk fields by \eqref{CFTtoAdS}. We thus claim that $\mu_\text{bulk}[\Phi_J=0]=0$ for $N<V$. Since $\Phi_J$ are the higher-spin fluctuations around empty AdS space, $\mu_\text{bulk}[\Phi_J=0]$ is the amplitude of empty AdS in quantum higher spin gravity, which should thus vanish in the continuum limit $V\rightarrow \infty$ at finite $N$. We can compute this observable directly at finite $V$ using \eqref{eq:W_def} and \eqref{eq:Z_phi} to get
\begin{equation}\label{eq:action_1}
\begin{split}
	\mu_\text{bulk}[\Phi_J=0] & = Z_\phi^{N} 
	\int d\Sigma_{i,j} e^{2\pi i  N \text{Tr}(\Sigma  G_0)} 
	\det(1+2\pi i \Sigma  G_0)^{-N}\,,
\end{split}
\end{equation}
where the integral $d\Sigma_{i,j}$ is over $V\times V$ hermitian matrices. We can then perform the change of variables $A=2\pi\Sigma G_0$ to get
\begin{equation}\label{eq:I0_result}
\begin{split}
	\mu_\text{bulk}[\Phi_J=0] & = Z_\phi^{N-V} \int dA_{i,j} \frac{e^{i \; N \text{Tr}(A)}}{\det(1+i A)^N}\\
	& = Z_\phi^{N-V}  \prod_{j=1}^V \int d\lambda_j\left(\prod_{i\ne j}^V (\lambda_i-\lambda_j) \right)  \left(\frac{e^{i \lambda_j}}{1+i \lambda_j}\right)^N\,,
\end{split}
\end{equation}
where in the second equation we used the Vandermonde determinant expression for hermitian matrices in terms of their eigenvalues $\lambda_j$. For each $\lambda_j$, we can then deform the contour around the upper half plane, and use the residue theorem successively around $\lambda_j = i$ to get
\begin{equation}
	\mu_\text{bulk}[\Phi_J=0] = Z_\phi^{N-V}  
	\left(\frac{2\pi \; e^{-N} N^{N-V}}{\Gamma(N)}\right)^V
	\prod_{k=2}^V (k\cdot(N-k+1))^{V-k+1}.
\end{equation}
The first term $\sim \det(G_0)^{N-V}$ is what we would naively get using the large $N$ bulk action in \cite{Aharony:2020omh} (up to overall factors from the $\Sigma,\phi_I$ integrals). The product term is a new contribution from the finite $N$ bulk action, that ensures that $\mu_\text{bulk}[\Phi_J=0] = 0$ for integer $N < V$.

\subsection{Critical theory}
\label{critical}

Finally, we can show how the discussion of the critical $U(N)$ theory in  \cite{Aharony:2020omh} for $2< d < 4$ generalizes to the finite $N$ theory we consider here. In the local formulation, the critical theory may be written as the low-energy limit of
\begin{equation} \label{eq:ctitical_deformation}
	 S_\text{crit}[\phi,\sigma] = \sum_{I=1}^N \int d^d x | \vec \nabla \phi_I(x) | ^2+   \frac{1}{2} \sigma(x) \sum_I \phi_I(x)^2 - \frac{N}{4\lambda} \sigma^2(x) \,,
\end{equation}
where $\sigma(x)$ is a local Hubbard-Stratonovich auxiliary field. In the bilocal formulation, this corresponds to (after regularization)
\begin{equation}
	S_\text{crit}[\eta,\Sigma, \sigma] = -2\pi i \; \text{Tr}( \Sigma\eta) + N  \left(-W[\Sigma] + \frac{1}{2} \sigma(x) \eta(x,x) - \frac{1}{4\lambda} \sigma^2(x)\right).
\end{equation}
In \cite{Aharony:2020omh} it was shown that we can take the $x_1\to x_2$ limit of the bilocal AdS/CFT map \eqref{CFTtoAdS} to get the off-shell relation
\begin{equation}\label{eq:st_rel}
	\eta(x,x) = \frac{1}{f_{d-2,0}} \lim_{\varepsilon \rightarrow 0} \varepsilon^{2-d} \Phi_0(x,\varepsilon)\,,
\end{equation}
which allows us to map the critical theory deformation to the bulk to get
\begin{equation} \label{eq:bulk_action_crit}
\begin{split}
	S_\text{bulk}[\Phi_J,\tilde\Phi_J, \sigma] &= -2\pi i  \sum_{J=0}^\infty \int \frac{d^d x dz}{z^{d+1}} \tilde\Phi_J(x,z) \Phi_J(x,z) \\&\quad+ N  \left(
	-W[\tilde\Phi_J]
	+ \frac{\sigma(x)}{2f_{d-2,0}} \lim_{\varepsilon \rightarrow 0} \varepsilon^{2-d} \Phi_0(x,\varepsilon) - \frac{1}{4\lambda} \sigma^2(x)
	\right)\,.
\end{split}
\end{equation}
For $2<d<4$, $\lambda$ is a relevant deformation, so we can flow to the IR by setting $\lambda=\infty$ directly. In the bulk action this eliminates the last term, so that integrating over $\sigma$ gives the off-shell relation $\varepsilon^{2-d}\Phi_0(x,\varepsilon) = 0$. This means that the boundary condition of $\Phi_0(x,z)$ at $z=0$ in the critical theory is now the same as the other $\Phi_J$'s, i.e. it starts (off-shell) at $z^\frac{d}{2}$ (rather than the $z^{d-2}$ behavior it has for the free CFT). So, as in the standard AdS/CFT correspondence, going to the critical theory is implemented just by changing the boundary condition for the bulk scalar field $\Phi_0$.

\section{Finite volume}
\label{finiteV}

In this section we derive the AdS dual of the free theory on a sphere, and in particular discuss the computation of the sphere free energy. We start by discussing how the AdS/CFT map discussed in the previous section for flat infinite space generalizes to the sphere $S^d$. We then use our finite $N$ formalism to schematically map the sphere free energy from the CFT to the bulk, and compare to previous computations in the literature. To make this map completely rigorous, we need to provide consistent regularizations in both the CFT and the bulk, which we discuss in the last subsection.

\subsection{AdS/CFT map on the sphere}

In the previous section, we discussed the AdS/CFT map that takes a field theory defined on $\mathbb{R}^d$ to a bulk theory defined on AdS with metric \eqref{AdSmet}, which corresponds to the upper half space $\mathbb{R}^d \times \mathbb{R}_+$ with a conformal boundary $\mathbb{R}^d$. We will now take the field theory to live on $S^d$, and map it to the $d+1$ dimensional hyperbolic space $\mathbb{H}^{d+1}$ with metric 
\es{sphereMet}{
ds_{\mathbb{H}^{d+1}}^{2}=dy^{2}+\sinh^{2}(y) d\Omega_{d}^{2}\,,
}
whose conformal boundary is $S^d$. Since $S^d$ is conformally equivalent to $\mathbb{R}^d$, the AdS spaces differ only by their boundary conditions. The AdS/CFT maps in Section \ref{adscftmap} can thus be adapted to the sphere by simply replacing all instances of the $\mathbb{R}^d$ metric, such as occur for integration measures (in the CFT or on the boundary of AdS) and lengths, by the $S^d$ metric, and replacing the 3-point functions and bulk-to-boundary propagators appearing in the equations by the same objects on $S^d$.

The only possible subtlety is the boundary conditions for $\eta(x_1,x_2)$ and $\Phi_J(x,z)$; we need to check if smooth configurations of the theory on the sphere obey the flat space boundary conditions we assumed, when we map them to flat space. The free field $\phi_I(\Omega)$ on an $S^d$ with unit radius is conformally related to $\mathbb{R}^d$ by the stereographic map
\begin{equation}
\phi_I\left(\Omega\right)\to\left(\frac{2}{1+|x|^{2}}\right)^{-\Delta_{0}}\phi_I\left(x\right)\,,
\end{equation}
where $\Omega$ is the spherical coordinate (and we left the dependence on the angular coordinates implicit). This implies that the second and third condition in the $\mathbb{R}^d$ boundary conditions \eqref{eq:bc_eta} are automatically satisfied for any field $\phi_I(\Omega)$ on $S^d$ that satisfies the first and fourth conditions. The boundary conditions for $\Phi_J(\Omega,z)$ on $\mathbb{H}^{d+1}$ follow from those of $\eta(x_1,x_2)$ just as in the flat space case, and so are the same as in Section \ref{adscftmap}. The conditions \eqref{eq:bc_eta} on the bi-local field in Section \ref{adscftmap} have their origin in normalizability with respect to a conformally-invariant inner product of bi-locals \cite{Dobrev:1977qv}. The CFT-to-AdS map takes this inner product to the standard $L_2$ inner product in the bulk. Therefore, the boundary conditions for any bulk manifold are simply those needed for $L_2$ normalizability there.

\subsection{Sphere free energy}

We will now use our sphere AdS/CFT map to relate the sphere free energy $F=-\log (Z)$ computed in the CFT to the same object computed in the bulk. For this it is useful to use the finite $N$ bilocal path integral \eqref{eq:eta_sigma_Z} in terms of $\eta$ and $\Sigma$. Since $\eta$ couples linearly to $\Sigma$, we can integrate $\eta$ out to get
\es{ZCFT}{
Z&= \int D\Sigma(x_1,x_2) e^{N  W[\Sigma]} \prod_{x_1,x_2}\delta(\Sigma(x_1,x_2)) = e^{N W[0]} =  Z_{\phi}^N\,,
}
where recall that $Z_\phi$ is the partition function of the free scalar local path integral \eqref{eq:Z_phi}, which depends on the regularization. Here, we wrote the delta function in terms of a $V$-dimensional lattice regularization introduced before. We can phrase this result in terms of the $\eta,\Sigma$ Feynman rules discussed in Section \ref{finiteN2}, by noting that there are simply no vacuum diagrams, since the quadratic action in \eqref{eq:S2_bi} has determinant $1$. So the answer for the sphere free energy in this formalism does not come from one-loop diagrams, but just from the constant first term on the last line of \eqref{eq:W_def}.

Note that in \eqref{ZCFT} we chose a specific normalization of the sphere free energy, which is the same as the one of $N$ free fields and is not affected by imposing the singlet condition. In general imposing the singlet condition could modify the free energy by multiplying it by a constant (for instance this happens if in $d=3$ it is imposed by coupling to a Chern-Simons gauge field at infinite level). In any computation of the sphere free energy one needs to specify the normalization chosen, and we discuss here how our particular normalization follows from the bulk actions we derived.

To compare the CFT free energy to the bulk, we need to be careful about the Jacobian factor that appears in deriving the bulk action from the CFT. As discussed in Section \ref{bulkAction}, if we choose the local bulk field redefinition parameters $f_{\Delta,J}$ and ${\tilde f}_{\Delta,J}$ as in \eqref{fnew}, then this Jacobian is 1, so all we need to do is compute the bulk action as written in \eqref{eq:bulk_Z}. Since the $\Phi_J$ couple linearly to the $\tilde\Phi_J$, we can integrate $\Phi_J$ out to get
\begin{equation}
	Z_\text{bulk} = \int D\tilde\Phi_J e^{N W_\text{bulk}[\tilde \Phi]} \prod_{J,x,z}\delta(\tilde \Phi_J(x,z)) = e^{N  W_\text{bulk}[0]} = Z_{\phi}^N\,,
\end{equation}
which exactly matches the CFT result. Again there is no contribution from loop diagrams. 
Note that the term $Z_{\phi}^N$ just comes from a constant term in the field theory $W$ that was directly copied to the bulk path integral, so it is defined here using the regularization of the CFT. If we want to write this constant term directly in a bulk language then we need to find a corresponding regularization in the bulk, and we will discuss this in the next subsection.

If we use the large $N$ formalism of \cite{Aharony:2020omh}, which is formally obtained by integrating out $\Sigma$ or $\tilde\Phi$ from the previous discussion, it is less clear how to reproduce the correct result for the sphere free energy.\footnote{See \cite{Jevicki:2014mfa} for calculations of the sphere free energy in the large $N$ bilocal CFT.} In this theory there would be contributions from one-loop and higher-loop diagrams in the bulk, and since we know the full answer still comes from a pre-factor sitting in front of the path integral, all these diagrams must cancel with contributions from the Jacobian, in a way that will depend on the details of the regularization.

It is interesting to contrast our calculation of the sphere free energy with previous computations in the literature. In our derivation, the non-zero contribution to the free energy $-N\log (Z_\phi)$ sits outside the path integral, which is why our calculation is immediate. Previous attempts to compute the free energy, starting with \cite{Giombi:2013fka} with many followups \cite{Giombi:2016pvg,Sun:2020ame,Giombi:2014iua,Gunaydin:2016amv,Brust:2016zns,Skvortsov:2017ldz,Basile:2018zoy,Basile:2018acb,Bae:2016rgm}, were hampered by the lack of an explicit action for the high-spin gravity theory in the bulk. In the absence of an action, the leading $O(N^1)$ large $N$ free energy could not be computed. However, the $O(N^0)$ 1-loop correction can still be derived if one assumes that the (off-shell) kinetic term in the bulk action takes the canonical two-derivative form for each spin, in which case the 1-loop correction can be computed by summing over the contributions of all physical and ghost modes that arise from a certain gauge fixing of this conjectured quadratic action. In the case of $N$ complex free scalars with $U(N)$ global symmetry, the resulting 1-loop correction was found to vanish as expected, while in the case of $N$ real free scalars with global symmetry $O(N)$, the 1-loop correction was found to be identical to the tree level term. This was justified by assuming that the dictionary $G_N\leftrightarrow1/N$ should be modified to $G_N\leftrightarrow1/(N+1)$ in this case. However, when the calculation was generalized to other examples of higher spin AdS/CFT in \cite{Giombi:2016pvg}, such as the duality of fermionic vector models in odd $d$, it was found that no such shift could explain the nonvanishing 1-loop free energy.

We now see that these previous calculations were unjustified in assuming that the off-shell quadratic action takes the canonical two-derivative term, as we see explicitly from our bulk action which has quartic-in-derivatives kinetic terms; the form of the off-shell action does not follow from just from the field content in the bulk. We thus see the ``match'' of the 1-loop correction for the $U(N)$ case as a coincidence, and the mismatch in all other cases as unsurprising. 

In the case of the critical vector models discussed in Section \ref{critical}, the sphere free energy will receive corrections order by order in $1/N$, which by construction will agree between the field theory and the gravity computations.

\subsection{Regularizations} \label{regs}

We now discuss how to consistently regularize the theories on either side of the duality. Note that in the finite $N$ theory of Section \ref{finiteN2} there are no loop diagrams, so a regularization is only needed for the vacuum amplitude discussed above; however, in the formalism of Section \ref{largeN} any computation must be regularized to make sense.

Our AdS/CFT mapping has been defined formally on the unregulated theories on either side of the duality, but to ensure that it really works, one must become convinced that the theories can somehow be consistently regularized. This means that a mapping must be found between  regulated, finite ``versions'' of either theory, which approaches the continuum mapping as the regulator is removed. 


On the field theory side, the CFT on the sphere is automatically IR regulated, and needs only to be UV regulated. One possible regulator which we discussed above is a lattice, but this breaks all the symmetries, so its analysis is very complicated. An alternative regulator is a hard cutoff on the eigenvalues of the Laplacian operator (acting on the original scalar fields) $|\nabla^2| <L^2$ (with $L\gg1$), which is sufficient to make the theory finite in the sense that the space of allowed bi-local fields is a finite-dimensional linear subspace of the unregulated one. Such a regulator breaks the conformal symmetry $SO(d+1,1)$ down to the sphere's isometries $SO(d+1)$, and we analyze how many modes remain in each $SO(d+1)$ representation in appendix \ref{reg1}.

In the bulk hyperbolic space \eqref{sphereMet} on the other hand, in order to get a finite number of modes one needs a UV regulator (say, a cutoff on the bulk Laplacian operator $|\nabla^2| \leq \Lambda^2$) as well as an IR regulator (say, boundary conditions at some finite radius $y=Y\gg1$), because the Laplacian has a continuous spectrum in hyperbolic space. These cutoffs can be chosen to be equal for every field of spin $J$ or to depend on the spin. The IR regulator breaks the bulk isometries $SO(d+1,1)\to SO(d+1)$, mirroring the conformal field theory. Again the regulators restrict the space of fields to a finite subspace. 

It is not easy to directly map the CFT regulator to the bulk or vice versa, since each regulator becomes very complicated when expressed in the other language. However, in order to construct a finite invertible mapping between the regulated theories on both sides, we do not necessarily need to have precisely the same regulator; it is enough that we have the same number of surviving modes in each $SO(d+1)$ representation. This is because the
map organizes itself into ``diagonal blocks'' corresponding to the surviving $SO(d+1)$ irreducible representations, and if each block is a square, a generic map like ours will be invertible with probability 1. Namely, we can either use the CFT-to-AdS mapping on the first line of \eqref{CFTtoAdS} on the modes that survive in the CFT, and its inverse to give us a regulated AdS-to-CFT mapping, or we can use the AdS-to-CFT mapping on the second line of \eqref{CFTtoAdS} on the modes that survive in the bulk, and its inverse will give us a regulated CFT-to-AdS mapping; the two options are different but both will reproduce the theory we want in the limit where the regulators are removed.

It can be argued (see appendix \ref{sec: dofs of reg theory})  that by taking the IR and UV regulators in the bulk to vary in some specific relation to the UV regulator of the CFT $Y(L),\Lambda(L)$, the degeneracy of each representation on the bulk side can be made to agree with the degeneracy in the CFT side.
The precise choice of regulators may be complicated, but it can be computed if necessary.
Making this choice leads to a precise mapping between the regulated theories on both sides (though not one that is conducive to explicit computations).


It should also be stated that there are some subtleties to IR regularization in the bulk -- specifically the choice of boundary conditions in the bulk -- due to the presence of high-spin fields in a prescribed gauge. The simplest boundary condition that makes sense is a Dirichlet condition on the completely parallel (to the boundary) and traceless (with respect to the boundary metric) components of the tensors, and it is sufficient to make the theory finite. However with this choice, somewhat inconveniently, the Laplacian is not self-adjoint (and the eigentensors correspondingly are not orthogonal) with respect to the standard inner product. It appears that no choice of boundary condition makes the Laplacian both self-adjoint and complete, but completeness appears to be sufficient to make the theory well-defined. The situation can be improved somewhat by changing the inner product into one with respect to which the Laplacian is both self-adjoint and complete, but this sacrifices ``locality'' in some sense, since the path integral cannot be interpreted as summing over sections of some vector bundle. Since the bulk action is not quite local to begin with, this isn't too much of a problem, and locality is restored as the cutoff is removed (the modified inner product approaches the standard one of hyperbolic space).

It's instructive to compare the theory under consideration with perturbative general relativity, which is a theory of a symmetric (but not, in general, traceless) rank-2 tensor. As discussed in \cite{Witten:2018lgb}, the choice of boundary conditions here is nontrivial, and ultimately involves, in one way or another, the conformal mode (the trace part) of the rank-2 tensor, which is absent in the higher-spin theory. We speculate however, that some fluctuations in the theory should perhaps be identified as ``trace-parts'' of others - they have the right scaling dimensions (in the sense of the CFT) and a ``wrong-sign'' kinetic term, much like the conformal mode in linearized general relativity. It is conceivable that boundary conditions that mix the high-spin fields with their trace-parts in some elaborate way could give a more well-behaved IR-regulated theory, but this investigation is left for future work. 

\section{Finite temperature}
\label{finiteT}

In this section we will discuss the bulk dual of the large $N$ vector model at finite temperatures $T$, namely the Euclidean theory on $M\times S^1_\beta$, for some $(d-1)$-dimensional manifold $M$, with $t \sim t + \beta$ (where $\beta = 1/T$ is the length of the $S^1$). This computation is more subtle than computations on flat space or on the sphere, because in order to impose the projection onto $U(N)$ singlets in the thermal partition function, we need to introduce a constant-in-space $U(N)$ holonomy $U = \exp(i \int_0^\beta A_t dt) \in U(N)$ on the thermal circle (coupled to the $\phi_I$) and to integrate over it. In a general dimension it is not clear how to introduce this holonomy in a local way without introducing additional degrees of freedom or flux sectors; for $d=3$ this can be done \cite{Giombi:2009wh} by coupling the vector models to a $U(N)_k$ Chern-Simons theory and taking $k\to \infty$. 

On general grounds, we expect in the large $N$ limit (with volume $V$) a phase transition between a low-temperature ``confined phase'' where low-dimension $U(N)$-singlets dominate the path integral, and a high-temperature ``deconfined phase'' whose thermodynamics can be approximated by ignoring the $U(N)$-singlet constraint. Schematically, we can think of the ``confined'' singlet degrees of freedom as the bi-local fields whose number scales as $V^2$ such that their free energy scales as $F \simeq T^{2d-1} V^2$, while the number of ``deconfined'' degrees of freedom scales as $N V$, so their free energy scales as $F \simeq N T^d V$. This suggests that the large $N$ phase transition should occur when $T^{d-1} V \simeq N$, and indeed this is what was found for the $d=3$ case on $S^2$ in \cite{Shenker:2011zf}. Both phases may be described by large $N$ saddle points in the integral over the holonomy. In the confined phase in the large $N$ limit the eigenvalue distribution of the saddle-point holonomy $U$ for $T^{d-1} V \ll N$ is approximately uniform, while in the deconfined phase the distribution is localized and all the eigenvalues approach one as $T^{d-1} V \gg N$.

%
For simplicity we discuss here the theory in infinite flat space. The discussion above makes it clear that the results in the large $N$ limit will depend on the order of limits between taking large $N$ and taking large $V$; if we take large $N$ first we will always be in the confined low-temperature phase, while if we take large $V$ first we will always be in the deconfined high-temperature phase. In any case in the large $N$ limit the path integral will be dominated by some specific background gauge field configuration $A_t^{cl}$ (with the other components of the gauge field vanishing), and at leading order in the saddle point approximation we can write it as
\begin{equation}
	Z(\beta) = \int D\phi_I \exp(-S[\phi_I,A_i^{cl}])
\end{equation}
with
\begin{equation}
	S[\phi_I,A_i^{cl}] = \int d^3 x (D_i \phi)_I^* (D^i \phi)_I,
\end{equation}
where $(D_i \phi)_I = ((\partial_i - i A_i^{cl}) \phi)_I$.

We begin by discussing the first case, where we take the large $N$ limit before taking the large volume limit. In this case we have a uniform eigenvalue distribution, which we can describe by choosing the gauge field matrix to take the value $(A_t^{cl})_{I,J} = \delta_{I,J} (\frac{2\pi I}{N \beta} + \alpha)$ for some constant $0 \leq \alpha < 2\pi / N \beta$. 
%
We would like to change variables to bilocal fields as in \cite{Aharony:2020omh}. The gauge-invariant bilocal operator is now
\begin{equation} \label{gibl}
\begin{split}
	G[\phi_I, A_i^{cl}](x_1,x_2)  & = \frac{1}{N} \phi^*(x_1) \exp\left(i \int_{x_2}^{x_1} A_i^{cl} dx^i\right) \phi(x_2) \\
	& = \frac{1}{N} \sum_{I=1}^N \phi_I^*(\vec{x}_1,t_1) \exp\left(i \left(\frac{2\pi I}{N\beta} + \alpha \right) (t_1-t_2) \right) \phi_I(\vec{x}_2,t_2).	
\end{split}
\end{equation}
In the second line we separated the space-time coordinates into spatial coordinates $\vec{x}$ and the time $t$, and substituted the classical value for $A_i$. Notice that the operator is invariant under the combined shift $t\mapsto t+\beta$ of the two coordinates, but not under a shift of a single coordinate since the holonomy is modified under such a shift; we have
\begin{equation} \label{period1}
	G(\vec x_1,t_1,\vec x_2, t_2) = G(\vec x_1,t_1+\beta,\vec x_2, t_2+\beta),
\end{equation}
while the periodicity in a single time coordinate is
\begin{equation} \label{period2}
G(\vec x_1,t_1,\vec x_2, t_2) = e^{-i \alpha N \beta} G(\vec x_1,t_1+N \ \beta,\vec x_2, t_2)=e^{i \alpha N \beta} G(\vec x_1,t_1,\vec x_2, t_2+N \ \beta).
\end{equation}

One way to perform computations in this theory is to compute the propagator of $\phi_I$ in the presence of the background gauge field and to plug it into all of our formulas. Alternatively we can swallow the gauge field $A^{cl}_i$ into a boundary condition of the local fields such that they obey ${\hat\phi}_I(\vec x,t+\beta) = e^{i (\frac{2\pi I}{N} + \alpha \beta)} {\hat\phi}_I(\vec x, t)$ and ${\hat\phi}_I^*(\vec x,t+\beta) = e^{-i (\frac{2\pi I}{N} + \alpha \beta)} {\hat\phi}^*_I(\vec x, t)$. 
In either approach we find that expectation value of the gauge-invariant bi-local \eqref{gibl}  is given by a sum over images of the zero-temperature propagator \eqref{saddle}
\begin{equation}\label{eq:thermal_2p_G}
	G_0(x_1,x_2,\beta) \equiv \langle G[\phi_I, A_i](x_1,x_2) \rangle =   \frac{1}{N} \sum_{I=1}^N \sum_{n=-\infty}^{\infty} G_0(\vec{x}_1, t_1 + n \beta, \vec{x}_2, t_2)
e^{-i n \beta (\frac{2\pi I}{N \beta} + \alpha)},
\end{equation}
consistent with the periodicities \eqref{period1}, \eqref{period2}. Note that for $d=3$ the sum over $n$ diverges, as expected since we expect the long-distance propagator at finite temperature to be the same as the propagator in $(d-1)$ dimensions which diverges in this case.

It is easy to perform the sum over $I$ in \eqref{eq:thermal_2p_G}, which localizes the sum to $n = N  m$ for integer $m$:
\begin{equation} \label{propn}
	G_0(x_1,x_2,\beta) = \sum_{m=-\infty}^{\infty} 
G_0(\vec{x}_1, t_1 + m N \beta, \vec{x}_2, t_2)
e^{-i m N \alpha \beta}.
\end{equation}
In the large $N$ limit all the images but $m=0$ die off, and we get back the zero temperature propagator $G_0(x_1,x_2)$. In principle we still need to integrate over $\alpha$, but clearly in the large $N$ limit this does not affect anything.

We see that in this phase we can think about the thermal circle as having effectively a length of $N \beta$, and so the large $N$ limit takes us back to $\mathbb{R}^d$. 
We can also see this by choosing a different form of the holonomy matrix; above we chose the holonomy to be $e^{i\alpha \beta}$ times the ``clock'' matrix, but by a gauge transformation we can alternatively choose the holonomy to be $e^{i\alpha \beta}$ times the ``shift'' matrix. With this choice, if we translate the holonomy into boundary conditions, we find ${\hat\phi}_{I+1}(\vec x,t+\beta) = e^{i \alpha \beta} {\hat\phi}_I(\vec x, t)$, such that all $N$ fields living on a circle of circumference $\beta$ can be described simply as a single field $\phi = \phi_1$ living on a circle of circumference $(N \beta)$, with $\phi(\vec x, t + N \beta) = e^{i N \alpha \beta} \phi(\vec x, t)$. In this language the bi-local can be written as $G(x_1,x_2) = \frac{1}{N} \sum_{I=1}^N \phi^*(\vec x_1, t_1 + I \beta) \phi(\vec x_2, t_2 + I \beta)$, and computing its expectation value directly gives \eqref{propn}.

Next we want to translate the finite temperature action to bi-local variables. This translation proceeds in exactly the same way as in sections \ref{largeN} or \ref{finiteN2}, just with the modified boundary conditions for $\phi_I$ corresponding to ``clock'' or ``shift'' matrices, which give the modified boundary conditions \eqref{period1} and \eqref{period2} for $G$ and for $\Sigma$ (with an opposite phase in the periodicity \eqref{period2} for $\Sigma$). In the large $N$ limit we just have the double-periodicity \eqref{period1}, and the classical saddle point $G_0$ is the same as for zero temperature. If we use the ``shift'' boundary conditions, the transformation from $\phi$ to $G$ of Section \ref{largeN}, or from $\phi$ to $\eta$ and $\Sigma$ of Section \ref{finiteN2}, proceeds in exactly the same way as in these sections, and we obtain precisely the same bi-local path integrals, just with the periodicity \eqref{period1}.

Next we wish to transform the bi-local path integral into the bulk. We can use precisely the same mapping that we used at zero temperature,
\begin{equation}\label{eq:inv_bulk_eta_map_thermal}
\begin{split}
	\Phi_{J}(\vec x, t, z) &=
	\int_{\gamma_J}\frac{d\Delta}{2\pi i}
	\int d^dy\, \int d^dx_1 d^dx_2\\
	&\times\frac12 \frac{f_{\Delta,J}}{N_{\Delta,J}}
	G_{\Delta,J}(\vec x,t,z|y)
	\thptild{\vec x_1,t_1}{\vec x_2,t_2}{\tilde\Delta,J}{y}\; \eta(\vec x_1,t_1,\vec x_2, t_2).
\end{split}
\end{equation}
Notice that since the integrand has a combined symmetry $t_1 \mapsto t_1 + \Delta t$, $t_2 \mapsto t_2 + \Delta t$, $t \mapsto t + \Delta t$, the resulting bulk field will have the property $\Phi_J(\vec x, t+\beta, z) = \Phi_J(\vec x, t, z)$, so the bulk fields now live on a thermal AdS space whose Euclidean time direction has periodicity $\beta$. Thus, as in other cases of the AdS/CFT correspondence, we find that the low-temperature phase of the CFT maps to a gravitational theory on thermal AdS space. Note that the same considerations should apply when the CFT is on a finite volume space such as $S^{d-1}$, if the zero temperature mapping on that space is known.

Above we described the mapping to the bulk only at leading order in the large $N$ limit. If we want to perform the mapping at finite $N$ there are various subtleties that need to be taken into account. First, the dominant holonomy is modified as a function of $T^{d-1} V / N$. More importantly, the fluctuations in the holonomy become important, and one has to integrate over different holonomies; it is not clear how to perform the mappings above for general holonomies, so we leave a discussion of this to future work. In addition, for finite $N$ we no longer obey the boundary condition $\lim_{|x_1|\rightarrow \infty}\eta(x_1,x_2)=0$ that we assumed (since $\eta$ is periodic in time with period $(N \beta)$), so one has to take this into account in the mapping (presumably by using the form of the mapping which is relevant for non-decaying fields, as discussed in \cite{Aharony:2020omh}).

We can also try to analyze the high-temperature phase, where we take the large volume limit before we take the large $N$ limit; in particular this is the relevant phase for the finite $N$ theory at large volume. On general grounds we expect this phase to map on the gravity side to a black hole background.
The saddle in this case is simply $A_t=0$ (with single valued $\phi_I$'s). Thus, in this phase the expectation value of the bi-local is simply
\begin{equation} \label{eq:thermal_2p_I_2}
	G_0(x_1,x_2, \beta) = \langle \phi_I^*(x_1) \phi_I(x_2) \rangle = \sum_{n=-\infty}^{\infty} G_0(\vec{x}_1, t_1 + n \beta, \vec{x}_2, t_2).
\end{equation}
This function, and also the variable $G(x_1,x_2)$ in this background field, is invariant to shifts by $\beta$ of $t_1$ and $t_2$ separately.
If we try to use the same mapping \eqref{eq:inv_bulk_eta_map_thermal} also in this background, it is unclear what constraint this extra periodicity gives on the bulk fields. Presumably, we need a different `black hole' geometry (as suggested also by the fact that the traces of the holonomy matrix are non-vanishing) and a different mapping in order to incorporate these periodicities in the bulk, and it would be interesting to understand how to do this.

\section*{Acknowledgements}

We would like to thank Jonathan Breuer, Steve Shenker and Douglas Stanford for useful discussions.
This work was supported in part  by an Israel Science Foundation center for excellence grant (grant number 2289/18), by grant no. 2018068 from the United States-Israel Binational Science Foundation (BSF), by the Minerva foundation with funding from the Federal German Ministry for Education and Research, by the German Research Foundation through a German-Israeli Project Cooperation (DIP) grant ``Holography and the Swampland'', and by a research grant from Martin Eisenstein. OA is the Samuel Sebba Professorial Chair of Pure and Applied Physics. SMC is supported by the Weizmann Senior Postdoctoral Fellowship.

\appendix

\section{Counting degrees of freedom in the regulated theory}\label{sec: dofs of reg theory}

In this appendix we count how many degrees of freedom remain in regulated versions of the mapping discussed in Section \ref{regs} that preserve an $SO(d+1)$ symmetry, in order to make it plausible that the regulators on both sides can be matched to give the same number of degrees of freedom in every $SO(d+1)$ representation.

\subsection{The field theory} \label{reg1}

On the CFT side, for the field theory on a sphere of radius 1, we can regulate the theory by keeping the modes of each scalar field $\phi_I$ that obey $\nabla^2 \leq  L(L+d-1)$ for some large $L$. Each scalar field then includes modes in the $(l,0,\cdots,0)$ representation of $SO(d+1)$ for $l=0,\cdots,L$. Then, for the bi-local fields we have the representations \cite{Dobrev:1977qv}:
\begin{equation}
\sum_{l_1,l_2=0}^L \left(l_{1},0,0,\dots\right)\otimes\left(l_{2},0,0,\dots\right) = \sum_{l_1,l_2=0}^L \sum_{i=0}^{\min\left(l_{1},l_{2}\right)}\sum_{j=0}^{\min\left(l_{1},l_{2}\right)-i}\left(l_{1}+l_{2}-2j-i,i,0,\dots\right).
\end{equation}
We can count how many times each $(l,m,0,\cdots,0)$ $SO(d+1)$ representation appears, for every $0 \leq m \leq l$ and whenever $l+m \leq 2L$. If $l < L$ the degeneracy is
\begin{equation}\label{eq:SO(d+1) irreps and degens of boundary theory}
d_{l,m} = (l-m+1)(L-l) + (\frac{l-m}{2}+1)^2 - \left[\frac{l-m}{2}\right]^2,
\end{equation}
while $[x]$ denotes the fractional part of $x$ (so the last term just subtracts $\frac{1}{4}$ whenever $(l-m)$ is odd). For $l \geq L$ (but still $l+m \leq 2L$) the corresponding equation is
\begin{equation}\label{eq:SO(d+1) irreps and degens of boundary theory large L}
d_{l,m} = (L-\frac{l+m}{2}+1)^2 - \left[\frac{l+m}{2}\right]^2.
\end{equation}

%

\subsection{The bulk} \label{reg2}

Next, we want to count how many representations remain in the bulk when we implement UV and IR regularizations as described in Section \ref{regs}.
In hyperbolic space, a rank $J$ symmetric traceless transverse tensor $\Phi_J$ can be decomposed \cite{Dobrev:1977qv} into plane wave normalizable orthogonal eigentensors $\Phi_{J}^{u,lm\omega}$ satisfying
\begin{equation}\label{eq: bulk quantum numbers}
\begin{array}{c}
-\nabla^{2}\Phi_{J}^{u,lm\omega}=\lambda_{u,J}\Phi_{J}^{u,lm\omega}\\
\lambda_{u,J}=u^{2}+\left(\frac{d}{2}\right)^{2}+J,
\end{array}
\end{equation}
where $l=J,J+1,\cdots$, $m=0,1,\cdots,J$, $\omega$ denotes the indices in the $(l,m,0,\cdots,0)$ representation of $SO(d+1)$, and $u$ is a continuous positive real parameter (the radial wavenumber of the solution). Each representation $(l,m,0,\cdots,0)$ appears from the $\Phi_J$'s with $J=m,m+1,\cdots,l$ (($l-m+1$) different fields), and its multiplicity depends on how many values of $u$ are kept by the regularization of $\Phi_J$.


After introducing an IR cutoff at some large radial position $y=Y$, with some boundary condition there, the allowed values of $u$ become discrete. We then introduce a bulk UV regulator $-\nabla^2<\Lambda^2$, which restricts for each field $u^2<u_{\rm max}^2 = \Lambda^2 -(d/2)^2-J$. We wish to relate these regulators to the parameter $L$ (the regulator in the field theory) in such a way that the multiplicity with which each representation appears, related to the number of allowed $u$ values, matches with that in the field theory. If we adjust the UV regulator separately for every value of $J$ and $l$, we have enough freedom to reproduce precisely the degeneracies \eqref{eq:SO(d+1) irreps and degens of boundary theory}, \eqref{eq:SO(d+1) irreps and degens of boundary theory large L}. Namely, we can first set the UV cutoffs to kill all modes with $l > 2L$. For $l=2L$ we want to keep only the $J=0$ mode and we fix $\Lambda$ so that for $m=0$ we obtain the correct degeneracy. Then, for $l=2L-1$ we want to keep only the modes of $J=0,1$, and we fix their degeneracies to match the degeneracies for $m=0,1$. And so on. 

Note that in the large $L$ limit the degeneracy of each fixed $(l,m,0,\cdots,0)$ mode in the field theory goes as $d_{l,m} \simeq (l-m+1)L$, so in this limit we just need each mode of $\Phi_J$ (with finite $J,l,m$) to have multiplicity $L$ (up to order one corrections). We can estimate the needed bulk cutoffs as follows.
%
In the bulk the maximal angular momentum is given roughly by $L_{bulk}^2 e^{-2Y}\approx u_{\rm max}^2$, coming from the centripetal term in the bulk Laplacian, and we need this $L_{bulk}$ to go as $L$. In addition, for $l$ of order 1, the eigentensors of the Laplacian are approximately plane waves and one can estimate that the number of allowed $u$ values is $\Lambda Y$, and we require this to be approximately $L$. Together these conditions give: 
\begin{equation}
\Lambda^{2} - L \approx L^{2}e^{-2Y}, \qquad L \approx \Lambda Y,
\end{equation}
which fixes the leading dependence of the regulators on $L$: $\Lambda,Y\propto\sqrt{L}$. Thus, as $L\to \infty$ the bulk regulators also go to infinity as expected.

\bibliographystyle{ssg}
\bibliography{finite_N_ref}

\begingroup\raggedright\begin{thebibliography}{10}

\bibitem{Maldacena:1997re}
J.~M. Maldacena, ``{The Large N limit of superconformal field theories and
  supergravity},'' {\em Int. J. Theor. Phys.} {\bf 38} (1999) 1113--1133,
  \href{https://arxiv.org/abs/hep-th/9711200}{{\tt hep-th/9711200}}. [Adv.
  Theor. Math. Phys.2,231(1998)].

\bibitem{Klebanov:2002ja}
I.~R. Klebanov and A.~M. Polyakov, ``{AdS dual of the critical O(N) vector
  model},'' {\em Phys. Lett.} {\bf B550} (2002) 213--219,
  \href{https://arxiv.org/abs/hep-th/0210114}{{\tt hep-th/0210114}}.

\bibitem{Aharony:2020omh}
O.~Aharony, S.~M. Chester, and E.~Y. Urbach, ``{A Derivation of AdS/CFT for
  Vector Models},'' \href{https://arxiv.org/abs/2011.06328}{{\tt 2011.06328}}.

\bibitem{deMelloKoch:2018ivk}
R.~de~Mello~Koch, A.~Jevicki, K.~Suzuki, and J.~Yoon, ``{AdS Maps and Diagrams
  of Bi-local Holography},'' {\em JHEP} {\bf 03} (2019) 133,
  \href{https://arxiv.org/abs/1810.02332}{{\tt 1810.02332}}.

\bibitem{Koch:2010cy}
R.~de~Mello~Koch, A.~Jevicki, K.~Jin, and J.~P. Rodrigues, ``{$AdS_4/CFT_3$
  Construction from Collective Fields},'' {\em Phys. Rev. D} {\bf 83} (2011)
  025006, \href{https://arxiv.org/abs/1008.0633}{{\tt 1008.0633}}.

\bibitem{Koch:2014aqa}
R.~de~Mello~Koch, A.~Jevicki, J.~a.~P. Rodrigues, and J.~Yoon, ``{Canonical
  Formulation of $O(N)$ Vector/Higher Spin Correspondence},'' {\em J. Phys. A}
  {\bf 48} (2015), no.~10 105403, \href{https://arxiv.org/abs/1408.4800}{{\tt
  1408.4800}}.

\bibitem{Vasiliev:1990en}
M.~A. Vasiliev, ``{Consistent equation for interacting gauge fields of all
  spins in (3+1)-dimensions},'' {\em Phys. Lett.} {\bf B243} (1990) 378--382.

\bibitem{Vasiliev:1992av}
M.~A. Vasiliev, ``{More on equations of motion for interacting massless fields
  of all spins in (3+1)-dimensions},'' {\em Phys. Lett. B} {\bf 285} (1992)
  225--234.

\bibitem{Vasiliev:1995dn}
M.~A. Vasiliev, ``{Higher spin gauge theories in four-dimensions,
  three-dimensions, and two-dimensions},'' {\em Int. J. Mod. Phys. D} {\bf 5}
  (1996) 763--797, \href{https://arxiv.org/abs/hep-th/9611024}{{\tt
  hep-th/9611024}}.

\bibitem{Boulanger:2015ova}
N.~Boulanger, P.~Kessel, E.~Skvortsov, and M.~Taronna, ``{Higher spin
  interactions in four-dimensions: Vasiliev versus Fronsdal},'' {\em J. Phys.
  A} {\bf 49} (2016), no.~9 095402,
  \href{https://arxiv.org/abs/1508.04139}{{\tt 1508.04139}}.

\bibitem{Sleight:2017pcz}
C.~Sleight and M.~Taronna, ``{Higher-Spin Gauge Theories and Bulk Locality},''
  {\em Phys. Rev. Lett.} {\bf 121} (2018), no.~17 171604,
  \href{https://arxiv.org/abs/1704.07859}{{\tt 1704.07859}}.

\bibitem{Bekaert:2016ezc}
X.~Bekaert, J.~Erdmenger, D.~Ponomarev, and C.~Sleight, ``{Bulk quartic
  vertices from boundary four-point correlators},'' in {\em {International
  Workshop on Higher Spin Gauge Theories}}, pp.~291--303, 2017.
\newblock \href{https://arxiv.org/abs/1602.08570}{{\tt 1602.08570}}.

\bibitem{Bekaert:2014cea}
X.~Bekaert, J.~Erdmenger, D.~Ponomarev, and C.~Sleight, ``{Towards holographic
  higher-spin interactions: Four-point functions and higher-spin exchange},''
  {\em JHEP} {\bf 03} (2015) 170, \href{https://arxiv.org/abs/1412.0016}{{\tt
  1412.0016}}.

\bibitem{Sleight:2016dba}
C.~Sleight and M.~Taronna, ``{Higher Spin Interactions from Conformal Field
  Theory: The Complete Cubic Couplings},'' {\em Phys. Rev. Lett.} {\bf 116}
  (2016), no.~18 181602, \href{https://arxiv.org/abs/1603.00022}{{\tt
  1603.00022}}.

\bibitem{Maldacena:2016hyu}
J.~Maldacena and D.~Stanford, ``{Remarks on the Sachdev-Ye-Kitaev model},''
  {\em Phys. Rev.} {\bf D94} (2016), no.~10 106002,
  \href{https://arxiv.org/abs/1604.07818}{{\tt 1604.07818}}.

\bibitem{Benedetti:2018goh}
D.~Benedetti and R.~Gurau, ``{2PI effective action for the SYK model and tensor
  field theories},'' {\em JHEP} {\bf 05} (2018) 156,
  \href{https://arxiv.org/abs/1802.05500}{{\tt 1802.05500}}.

\bibitem{Giombi:2013fka}
S.~Giombi and I.~R. Klebanov, ``{One Loop Tests of Higher Spin AdS/CFT},'' {\em
  JHEP} {\bf 12} (2013) 068, \href{https://arxiv.org/abs/1308.2337}{{\tt
  1308.2337}}.

\bibitem{Jevicki:1979mb}
A.~Jevicki and B.~Sakita, ``{The Quantum Collective Field Method and Its
  Application to the Planar Limit},'' {\em Nucl. Phys.} {\bf B165} (1980) 511.

\bibitem{Dobrev:1977qv}
V.~K. Dobrev, G.~Mack, V.~B. Petkova, S.~G. Petrova, and I.~T. Todorov,
  ``{Harmonic Analysis on the n-Dimensional Lorentz Group and Its Application
  to Conformal Quantum Field Theory},'' {\em Lect. Notes Phys.} {\bf 63} (1977)
  1--280.

\bibitem{Aharony:2021ovo}
O.~Aharony, S.~M. Chester, and E.~Y. Urbach, ``{AdS from CFT for scalar QED},''
  {\em Phys. Rev. D} {\bf 104} (2021), no.~12 126011,
  \href{https://arxiv.org/abs/2109.05512}{{\tt 2109.05512}}.

\bibitem{Maldacena:2011jn}
J.~Maldacena and A.~Zhiboedov, ``{Constraining Conformal Field Theories with A
  Higher Spin Symmetry},'' {\em J. Phys.} {\bf A46} (2013) 214011,
  \href{https://arxiv.org/abs/1112.1016}{{\tt 1112.1016}}.

\bibitem{Jevicki:2014mfa}
A.~Jevicki, K.~Jin, and J.~Yoon, ``{1/N and loop corrections in higher spin
  AdS$_4$/CFT$_3$ duality},'' {\em Phys. Rev. D} {\bf 89} (2014), no.~8 085039,
  \href{https://arxiv.org/abs/1401.3318}{{\tt 1401.3318}}.

\bibitem{Giombi:2016pvg}
S.~Giombi, I.~R. Klebanov, and Z.~M. Tan, ``{The ABC of Higher-Spin AdS/CFT},''
  {\em Universe} {\bf 4} (2018), no.~1 18,
  \href{https://arxiv.org/abs/1608.07611}{{\tt 1608.07611}}.

\bibitem{Sun:2020ame}
Z.~Sun, ``{AdS one-loop partition functions from bulk and edge characters},''
  {\em JHEP} {\bf 12} (2021) 064, \href{https://arxiv.org/abs/2010.15826}{{\tt
  2010.15826}}.

\bibitem{Giombi:2014iua}
S.~Giombi, I.~R. Klebanov, and B.~R. Safdi, ``{Higher Spin AdS$_{d+1}$/CFT$_d$
  at One Loop},'' {\em Phys. Rev.} {\bf D89} (2014), no.~8 084004,
  \href{https://arxiv.org/abs/1401.0825}{{\tt 1401.0825}}.

\bibitem{Gunaydin:2016amv}
M.~G\"unaydin, E.~D. Skvortsov, and T.~Tran, ``{Exceptional $F(4)$ higher-spin
  theory in AdS$_{6}$ at one-loop and other tests of duality},'' {\em JHEP}
  {\bf 11} (2016) 168, \href{https://arxiv.org/abs/1608.07582}{{\tt
  1608.07582}}.

\bibitem{Brust:2016zns}
C.~Brust and K.~Hinterbichler, ``{Partially Massless Higher-Spin Theory},''
  {\em JHEP} {\bf 02} (2017) 086, \href{https://arxiv.org/abs/1610.08510}{{\tt
  1610.08510}}.

\bibitem{Skvortsov:2017ldz}
E.~D. Skvortsov and T.~Tran, ``{AdS/CFT in Fractional Dimension and Higher Spin
  Gravity at One Loop},'' {\em Universe} {\bf 3} (2017), no.~3 61,
  \href{https://arxiv.org/abs/1707.00758}{{\tt 1707.00758}}.

\bibitem{Basile:2018zoy}
T.~Basile, E.~Joung, S.~Lal, and W.~Li, ``{Character Integral Representation of
  Zeta function in AdS$_{d+1}$: I. Derivation of the general formula},'' {\em
  JHEP} {\bf 10} (2018) 091, \href{https://arxiv.org/abs/1805.05646}{{\tt
  1805.05646}}.

\bibitem{Basile:2018acb}
T.~Basile, E.~Joung, S.~Lal, and W.~Li, ``{Character integral representation of
  zeta function in AdS$_{d+1}$. Part II. Application to partially-massless
  higher-spin gravities},'' {\em JHEP} {\bf 07} (2018) 132,
  \href{https://arxiv.org/abs/1805.10092}{{\tt 1805.10092}}.

\bibitem{Bae:2016rgm}
J.-B. Bae, E.~Joung, and S.~Lal, ``{One-loop test of free SU(N ) adjoint model
  holography},'' {\em JHEP} {\bf 04} (2016) 061,
  \href{https://arxiv.org/abs/1603.05387}{{\tt 1603.05387}}.

\bibitem{Witten:2018lgb}
E.~Witten, ``{A note on boundary conditions in Euclidean gravity},'' {\em Rev.
  Math. Phys.} {\bf 33} (2021), no.~10 2140004,
  \href{https://arxiv.org/abs/1805.11559}{{\tt 1805.11559}}.

\bibitem{Giombi:2009wh}
S.~Giombi and X.~Yin, ``{Higher Spin Gauge Theory and Holography: The
  Three-Point Functions},'' {\em JHEP} {\bf 09} (2010) 115,
  \href{https://arxiv.org/abs/0912.3462}{{\tt 0912.3462}}.

\bibitem{Shenker:2011zf}
S.~H. Shenker and X.~Yin, ``{Vector Models in the Singlet Sector at Finite
  Temperature},'' \href{https://arxiv.org/abs/1109.3519}{{\tt 1109.3519}}.

\end{thebibliography}\endgroup
\end{document}